\documentclass[letter]{emulateapj}

\begin{document}
\title{The SDSS Coadd: Cross-Correlation Weak Lensing and Tomography of Galaxy Clusters}

\author{Melanie Simet$^{1,2}$, Jeffrey M. Kubo$^3$, Scott Dodelson $^{1,2,3}$, James T. Annis$^3$, Jiangang Hao$^3$, David Johnston$^3$, Huan Lin$^3$, Ribamar R. R. Reis$^4$, Marcelle Soares-Santos$^3$, Hee-Jong Seo$^{5}$}
\affil{$^1$Department of Astronomy \& Astrophysics, The
University of Chicago, Chicago, IL~~60637}
\affil{$^2$Kavli Institute for Cosmological Physics, Chicago, IL~~60637}
\affil{$^3$Center for Particle Astrophysics, Fermi National
Accelerator Laboratory, Batavia, IL~~60510}
\affil{$^4$Instituto de F\'{i}sica, Universidade Federal do Rio de Janeiro, Brazil}
\affil{$^5$Berkeley Center for Cosmological Physics \& Berkeley Lab, University of California, Berkeley, CA~~94720}

 \begin{abstract}
The shapes of distant galaxies are sheared by intervening galaxy clusters. We examine this effect in Stripe 82, a 275 square degree region observed multiple times in the Sloan Digital Sky Survey and coadded to achieve greater depth.  We obtain a mass-richness calibration that is similar to other SDSS analyses, demonstrating that the coaddition process did not adversely affect the lensing signal.  We also propose a new parameterization of the effect of tomography on the cluster lensing signal which does not require binning in redshift, and we show that using this parameterization we can detect tomography for stacked clusters at varying redshifts.  Finally, due to the sensitivity of the tomographic detection to accurately marginalizing over the effect of the cluster mass, we show that tomography at low redshift (where dependence on exact cosmological models is weak) can be used to constrain mass profiles in clusters.
\end{abstract}


\section{Introduction}

Weak lensing, the distortion of observed galaxy shapes by the gravitational potential of large-scale structure, has great potential to help determine cosmological parameters \citep{TaskForce, HoekstraJain, Huterer, Mellier, Munshi, Peacock}.  Lensing by galaxy clusters can constrain the mass and mass distribution of those clusters \citep{KaiserSquires, SchneiderBook, Johnston}, which has implications for the amplitude of the matter power spectrum $\sigma_8$, the matter density $\Omega_m$, and the evolution of dark energy \citep{Marian, Wang, TaskForce}.  

Since the lensing cross-section varies with the distances between the observer, the lens, and the lensed galaxy, it is also possible to obtain information about the time-dependent cosmic geometry.  Tomography can constrain parameters such as the dark energy density $\Omega_{\Lambda}$ and its equation of state $w$ \citep{Hu} and can also test general relativity on large scales \citep{Zhao}.  Tomographic analysis for cosmology requires deep, wide, high-resolution surveys \citep{Bernstein, TaskForce, Peacock}, so there is great potential for results from upcoming large-scale surveys such as the Dark Energy Survey (DES; {\tt http://www.darkenergysurvey.org}).

Tomography has previously been observed around a small number of clusters.  The change in shear with redshift can be observed by binning source galaxies into redshift slices and determining the amplitude of the signal in each bin around single clusters, as \citet{Medezinski} and \citet{Taylor} have done; in addition \citeauthor{Taylor} fit a three-dimensional gravitational potential for the clusters in their survey.  \citet{Gavazzi} and \citet{Shan} take an inverse approach, using the expected change with redshift to infer the redshift of potential clusters identified through shear peaks and to distinguish noise peaks from real clusters.  \citet{Simon} look for shear peaks in three-dimensional convergence maps of the Abell 901/902 supercluster, detecting additional structure behind the known clusters using the lensing strength at a series of redshift slices.  Here we restrict ourselves to the question of the redshift-distance relation, using a stacked sample of many clusters rather than making a detection for single clusters in high signal-to-noise data.

In this work, we detect lensing around the clusters in Stripe 82 of the Sloan Digital Sky Survey (SDSS), a region observed multiple times so that it probes $\sim2$ magnitudes deeper than the SDSS sample overall, reaching a depth (50\% completeness) of 23 in the $i$-band \citep{SDSS, Stripe82, Annis}.  Like the DES, Stripe 82 of the SDSS achieves its depth by coadding many images of the same region of the sky.  This process can introduce systematic errors to the lensing signal (\citealt[Part 3, \S 3.3]{SchneiderBook}), so one motivation for this study is to check if coadding adversely affects the cluster lensing signal.  The deeper sample also opens up the possibility of detecting tomography. In principle, tomography offers the promise of determining cosmological parameters, but in this study we aim only to detect the greater shear in distant galaxies.

Our lensing and cluster data are described in \S \ref{Data}.  We analyze the data using a likelihood method described in \S \ref{LMethod}.  Results of the analysis are given in \S \ref{Results}.  

\section{Data}\label{Data}

\subsection{Images}

For our lensing catalog, we use the publicly available coadded images from Stripe 82 \citep{Annis}.  The total area of the stripe is $~275$ deg$^2$, with a declination between -1.25$^{\circ}$ and +1.25$^{\circ}$ and a right ascension from 20\textsuperscript{h} to 4\textsuperscript{h}. The stripe was observed a number of times, in some cases with the same seeing requirements as the rest of the SDSS and in others as part of a supernova survey that also took images in worse seeing and during times of bright moonlight or bad photometric conditions.  After cuts were made, the best images were coadded to obtain deeper images (20-30 images combined for most of the stripe, selected based on seeing in the $r$ band, $r$-band sky brightness, and sufficiently small photometric corrections required during photometric calibration); the median seeing in the coadded images is 1.1''.  See \citet{Annis} and \S 3 of \citet{coadd} for more detail.  The \citet{Johnston} sample, in comparison, covers single (not coadded) images in a subset of Data Release 4 selected for extinction and distance from the survey edge, some of which overlap with our sample \citep{SheldonStripe82}.

\begin{figure}
  \begin{center}
  \begin{tabular}{c}
  \includegraphics[width=0.45\textwidth]{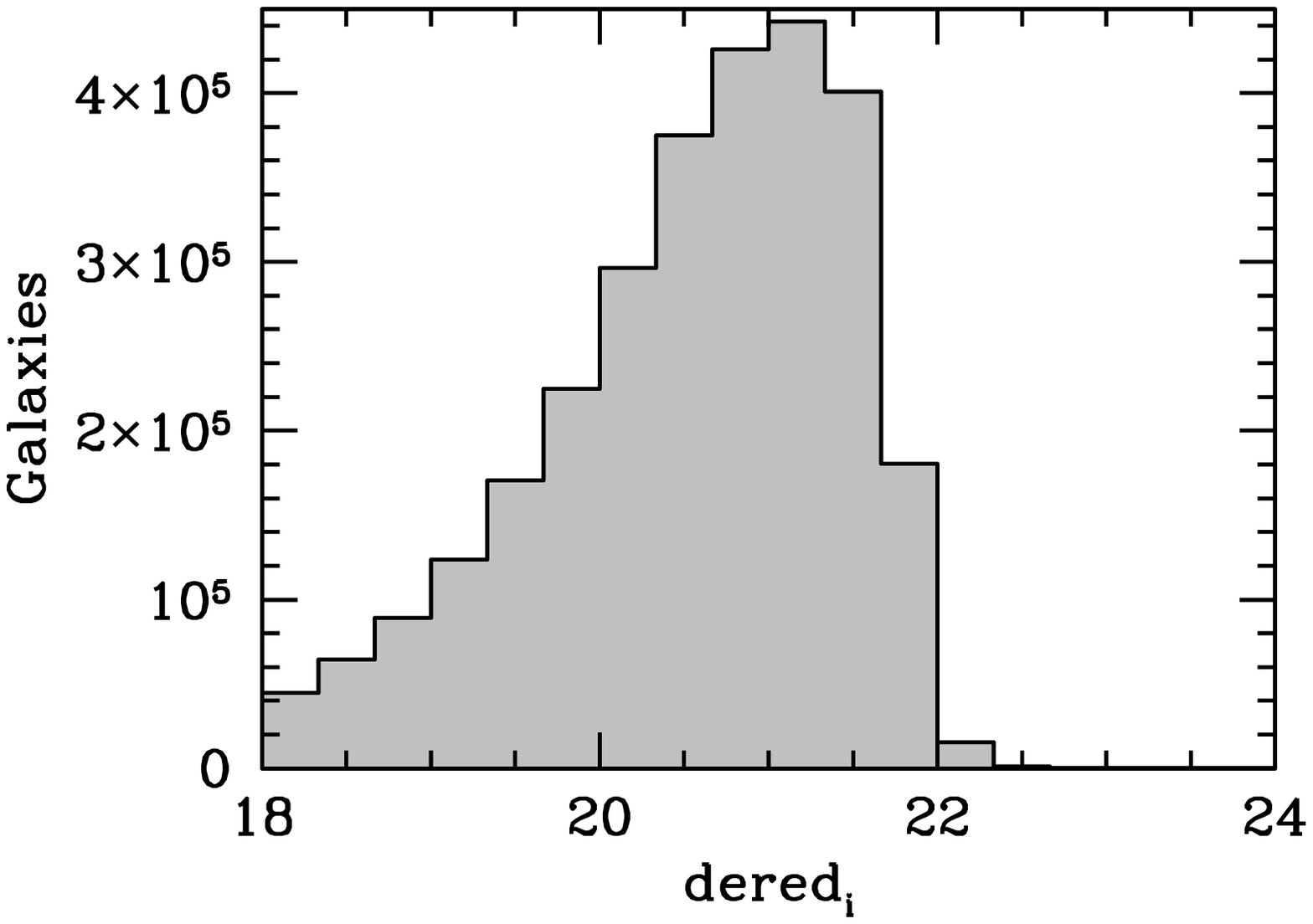} \\
  \includegraphics[width=0.45\textwidth]{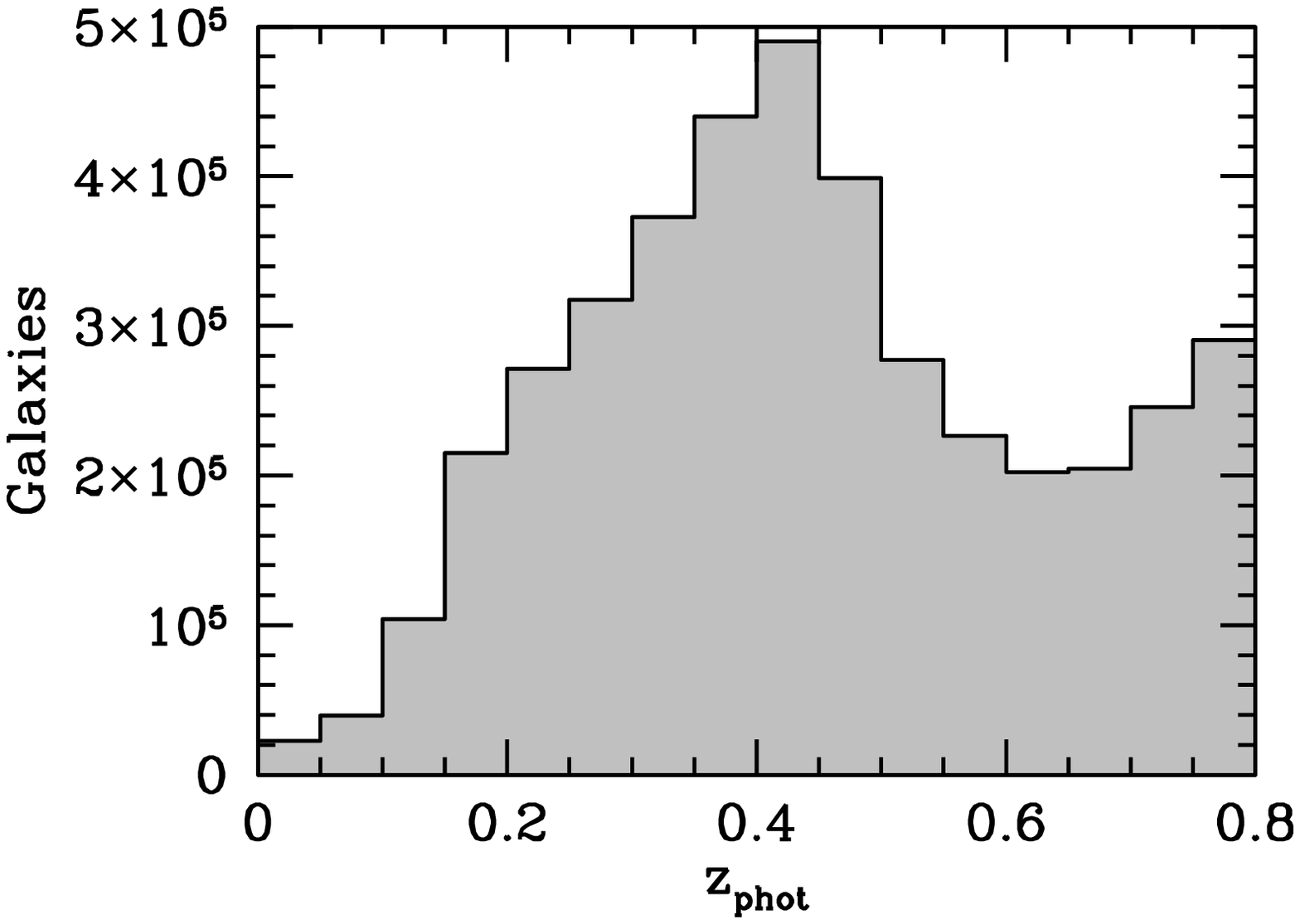}
  \end{tabular}
  \caption{Histograms of i-band magnitudes (top) and photometric redshifts (bottom) for our selected objects in the Stripe 82 coadd.  While our selection criteria allow i-band magnitudes up to 24, most of the galaxies with i-band magnitudes above 22 have been removed by our photo-z selection criteria; the $\sigma_z$ selection criteria also removed many galaxies beyond $z=0.5$.}\label{iband-histogram}
  \end{center}
\end{figure}

We select objects from the Stripe 82 images that are classified as having an object type of 3, which are objects the pipeline identified as galaxies. The galaxies are then selected for extinction-corrected i-band magnitudes in the range $18.0<dered_{i}<24.0$ and for object size ($M_{rrcc}$, the sum of the second-order moments in the detector row and column directions) greater than 1.5 times the size of the best-fit point spread function ($M_{rrcc}^{psf}$) at the position of each galaxy.  We also reject objects that contain saturated pixels or that triggered flags indicating problems in the adaptive moment measurements.  To correct for the PSF anisotropy and dilution, we use the linear PSF correction scheme described in Appendix B of \citet{HirataSeljak}.  We fit polynomials to the residual differences between measured and model PSF ellipticities and sizes for bright stars, in order to improve the PSF model subsequently used in correcting the measured galaxy shapes.  In addition, we find that the average ellipticities in each camera column (detector) are nonzero, so we subtract this small bias from the final results to force the averages to zero.  Further details are available in \cite{cosmicshear}.  We reject galaxies with a total corrected ellipticity $e_{corr}>1.4$. A histogram of the magnitudes of the galaxies included in our coadded catalog, after the photo-z cuts described in \S \ref{photozcut}, is shown in Fig. \ref{iband-histogram}.  

We select clusters from the MaxBCG catalog of \citet{Koester}.  The MaxBCG algorithm detects galaxy clusters by matching the galaxy distribution to a cluster model that depends on the clustering in spatial and color space as well as the presence of a brightest cluster galaxy (BCG) at the center of the cluster. Cluster member galaxies are selected as within $\pm 2\sigma$ of the corresponding red sequence, fainter than the identified BCG but brighter than $0.4L_{*}$ at the cluster redshift.  The catalog covers the redshift range $0.1<z<0.3$.  We then divide the catalog into richness bins, where the richness $N_{200}$ counts the number of cluster member galaxies within a radius $r_{200}$ of the cluster center. The radius $r_{200}$ is the radius inside which the average density of the cluster is 200 times the critical density $\rho_c$ of the universe at the lens redshift; in the case of the MaxBCG catalog, an observational proxy for the theoretical $r_{200}$ is used, but previous studies have found the two to be in good agreement \citep[e.g.][]{Johnston}.  There are 492 clusters with richness $N_{200} \ge 10$ with a maximum richness of 88.  We bin these clusters into the 6 richness bins shown in in Table \ref{richtable}, which follows the scheme of \citet{Johnston} to facilitate comparison.  We note that the richest bin has only one cluster, but we confirm in \S \ref{Results} that this does not bias our results. The \citeauthor{Johnston} DR4 sample is much brighter, so the background galaxies are at lower redshift and are less dense than our Stripe 82 sample (at about 1 galaxy per square arcminute in the DR4 sample compared to 6 galaxies per square arcminute in the coadd). On the other hand, the selected region of DR4 covers much more area than Stripe 82 and so it contains a larger number of clusters in our redshift range.  The combination means our study is complementary to \citet{Johnston} and our constraints on mass as a function of richness should be consistent with their results.

\begin{table}
\begin{center}
\begin{tabular}{|c|c|}
\hline
$N_{200}$ & Number of clusters \\
\hline
10-11 & 167 \\
12-17 & 182 \\
18-25 & 85 \\
26-40 & 38 \\
41-70 & 19 \\
88 & 1 \\
\hline
\end{tabular}
\label{richtable}
\caption{Richness ranges for our selected bins}
\end{center}
\end{table}

\subsection{Photometric redshift catalog}\label{photozcut}

In our analysis we use photometric redshifts for the SDSS coadd catalog generated by \citet{photoz}.  Briefly, this method uses the same neural network algorithm that was used in the SDSS DR6 \citep{DR6, DR6-photoz}.  Here a spectroscopic training sample of 83,000 galaxies from five different surveys are used: the SDSS DR7 \citep{coadd}, CNOC2 \citep{CNOC2}, WiggleZ \citep{WiggleZ}, DEEP2 \citep{DEEP2}, and VVDS \citep{VVDS}.  Futher details on the SDSS coadd photometric redshift catalog can be found in \citet{photoz}.  

Cuts were made on the photometric redshifts to select galaxies with $z_{phot}<0.8$ and photometric redshift error $z_{err}<0.1$.  After these cuts, 4.12 million background galaxies remain with projected distance less than 5 h$^{-1}$ Mpc from a cluster center.  The photometric redshift distribution of these galaxies is shown in Fig.~\ref{iband-histogram}.

\section{Methodology}\label{LMethod}

\subsection{Theoretical Background}

The azimuthally averaged tangential shear signal around any distribution of mass with projected density $\Sigma(r)$ is given at radius $R$ by
\begin{equation}
\langle\gamma_t\rangle(R)=\frac{\Delta\Sigma_t(R)}{\Sigma_c} \equiv \frac{\bar{\Sigma}(R) - \langle\Sigma(R)\rangle}{\Sigma_c}
\end{equation}
where brackets indicate the azimuthal average, the bar indicates an average over all radii interior to $R$, and $\Sigma_c$ is defined by
\begin{equation}
 \Sigma_c=\frac{c^2}{4\pi G} \frac{D_s}{D_d D_{ds}}
\end{equation}
where $D_s$, $D_d$ and $D_{ds}$ are the angular diameter distances from the observer to the source, from the observer to the lens, and from the lens to the source, respectively \citep{SchneiderBook}.  The most direct view of the lensing signal in the data is the tangential $\Delta\Sigma$ profile,
\begin{equation} 
\Delta\Sigma_t(R) = \Sigma_c \langle\gamma_t\rangle(R),
\end{equation}
which removes the dependence on background galaxy redshift and most of the dependence on cluster redshift.  We can also plot the  $\Delta\Sigma_x$ profile, using $\gamma_x$, the component of the shear oriented at 45$^{\circ}$ to a vector from the object to the center of the lens, instead of $\gamma_t$, the component oriented at 90$^\circ$.  The $\Delta\Sigma_x$ profile should be zero.

We assume that each cluster has a Navarro-Frenk-White profile \citep[][NFW]{NFW} with concentration $c_{200}=4$ \citep{Concentration, Johnston} and one free parameter $M_{200}$, defined in a similar way to $N_{200}$ as the mass inside a sphere of radius $r_{200}$.  The density profile is 
\begin{equation}
\rho(r)=\frac{\delta_{c} \rho_c}{(r/r_{s})(1+r/r_{s})^2}
 \end{equation}
with scale radius 
\begin{equation}
 r_s \equiv \frac{r_{200}}{c_{200}}=\frac{1}{c_{200}}\left(\frac{3M_{200}}{800\pi \rho_c}\right)^{1/3},
  \end{equation}
and
\begin{equation} 
\delta_{c} \equiv \frac{200}{3} \frac{c_{200}^3}{\ln(1+c_{200}) - c_{200}/(1+c_{200})}.
 \end{equation}
The tangential shear is then
\begin{equation}
\label{nfwgamma}
\gamma_t = \frac{r_s\delta_{c}\rho_c}{\Sigma_c}g(R/r_s)
\end{equation}
where $g(x)$ resembles a smoothed broken power law, steeper than $1/r$ at radii $r>r_s$ and shallower, asymptotically approaching 1, at $r<r_s$.  The full form can be found in \citet{Wright}.  

There are several sources of contamination that must be accounted for to extract an unbiased estimate of $M_{200}$ for a cluster or a stacked set of clusters in a given richness bin.  At very small distances from the cluster center ($\sim 0.1 h^{-1}$ Mpc in the lens plane) a term from the mass in the brightest cluster galaxy (BCG) becomes important, and at very large radii ($\sim 5-10 h^{-1}$ Mpc in the lens plane) the overlapping signal from nearby haloes comes to dominate.  To avoid such effects, we limit our analysis to the background galaxies in the annulus $0.1$-$5 h^{-1}$ Mpc in the lens plane from the cluster center.  Intervening large-scale structure also has an effect on the observed strength of the lensing; however, combining clusters at different points on the sky should remove the dependence on the large-scale structure as the extra lensing along the different lines of sight should be uncorrelated \citep{Johnston}.  There is also a non-negligible chance that the selected BCG is not at the center of the cluster, which would dilute the shear signal at smaller radii by essentially averaging over an annulus that includes background galaxies at varying radii from the cluster center.  We seek to understand this correction with mock catalogs as described in \S \ref{Mocks}.  

A further distortion to the shear profile comes from the misidentification of galaxies in the cluster as galaxies behind the cluster, which dilutes the average shear signal.  The correction for this effect is found by comparing the (appropriately weighted) number density of galaxies close to clusters with the corresponding number density around random points on the sky, as described in \citet[\S 4.1]{Sheldon}.  Specifically, for background galaxies $i$ around a set of clusters $j$ and for background galaxies $k$ around a set of randomly chosen points $l$, the correction factor is given by
\begin{equation}
  C(r) = \frac{N_{random}}{N_{clusters}}\frac{\sum_{i,j} \langle \Sigma_{c,ij}^{-1} \rangle ^2/(\sigma_{\gamma}^{i})^{2} }  {\sum_{k,l} \langle \Sigma_{c,kl}^{-1} \rangle^2/(\sigma_{\gamma}^{k})^{2}}  
\end{equation}
where $\sigma_{\gamma}$ is the error from both the intrinsic ellipticity error and the measurement error, $N_{random}$ and $N_{clusters}$ are the number of random points and clusters respectively, and the expectation value $\langle \Sigma_c^{-1} \rangle$ is taken to make the function well-behaved near $D_s=D_d$.  This correction does not account for magnification, but the error induced in the correction by magnification is of the same order as the convergence, itself the same order as the shear \citep{magnification} which is $\approx 0.01$, and therefore we expect that it will be lower than other sources of noise in this analysis.  The spread of photo-z uncertainty is accounted for in the calculation of $\langle \Sigma_c^{-1} \rangle$.  

\begin{figure} 
  \begin{center}
  \includegraphics[width=0.45\textwidth]{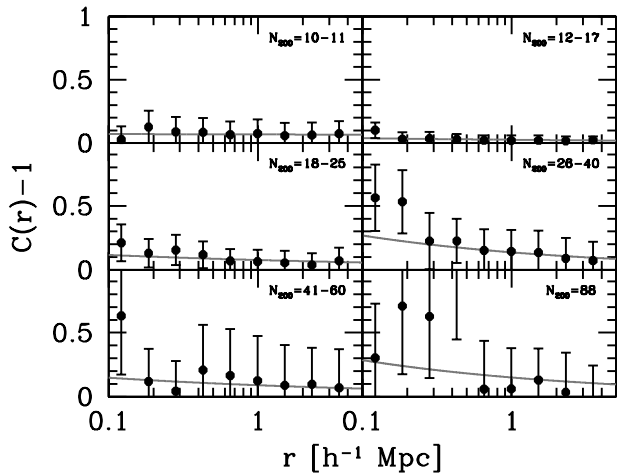}
    \caption{Correction factor due to misidentification of cluster member galaxies.  Best-fit powerlaws are shown.}
    \label{boost}
  \end{center}
\end{figure}

When performing the mass fits, the individual shears are multiplied by $C(r)$ to remove the clustering bias.  We plot the correction for the different richness bins of the coadd in Fig. \ref{boost}.  We find that the correction scales approximately as a powerlaw in radius, with indices from $~0.2-1.4$. 

After applying these corrections to the underlying shear profiles, we bin the clusters in richness and perform a $\chi^2$ fit to the data in each bin to obtain an estimate of $M_{200}$ in the given bin.  We have fixed the concentration to $c_{200}=4$.  We expect that the masses will be related to the richnesses via a power law,
\begin{equation}\label{mass-richness}
 M_{200} = M_{200|20}\left(\frac{N_{200}}{20}\right)^{\alpha}
  \end{equation}
where $M_{200|20}$ is the mass of a cluster with $N_{200}=20$.  When doing the fits within each bin, we marginalize over the powerlaw index $\alpha$ to avoid biasing the masses within each bin.  We then fit a relation of this form to the six masses we have obtained, one for each bin.

\subsection{Tomography}

The second part of our analysis is to search for a tomographic signal. We write $\Sigma_c$ for a source redshift $z$ as
\begin{equation}
  \frac{1}{\Sigma_c(z)} =  \frac{1}{\Sigma_c(\bar{z})}\left(1+a\left(\frac{\Sigma_c(\bar{z})}{\Sigma_c(z)} -1\right)\right) 
\end{equation}
where $\bar z=0.45$ is the median redshift of the background galaxies and $a$ is a free parameter whose true value is one.  We assume a fiducial cosmology of $\Omega_m=0.3$, $\Omega_{\Lambda}=0.7$, and $H_0=100h$ km/s/Mpc to generate $\Sigma_c$.  A survey with no sensitivity to the background redshifts would not be able to constrain $a$ whereas a deep survey with reliable redshifts would place tight constraints on $a$. For this part of the analysis, we vary $M_{200|20}$, fixing the slope $\alpha$ to its best fit value. Then we marginalize over the mass scaling to obtain a likelihood for $a$. Ultimately, one would like to use tomography to constrain the geometry of the Universe, but mock catalogs suggest that Stripe 82 will not have the sensitivity to probe cosmological parameters, so we report the more agnostic constraint on $a$.

For a more visual representation of the tomographic signal, we bin the background galaxies by redshift, then determine the amplitude of the NFW shear signal (not $\Delta\Sigma$ signal) for each bin, using all the clusters scaled together by $N_{200}$ as in the previous section.  We do not correct for the differing redshifts of the clusters, however.  We then plot the shear predicted by our fit for a galaxy located at 1 $h^{-1}$ Mpc from a $N_{200}=20$ cluster for each redshift bin.

\subsection{Mock catalogs}\label{Mocks}

\begin{figure} 
  \begin{center}
  \includegraphics[width=0.45\textwidth]{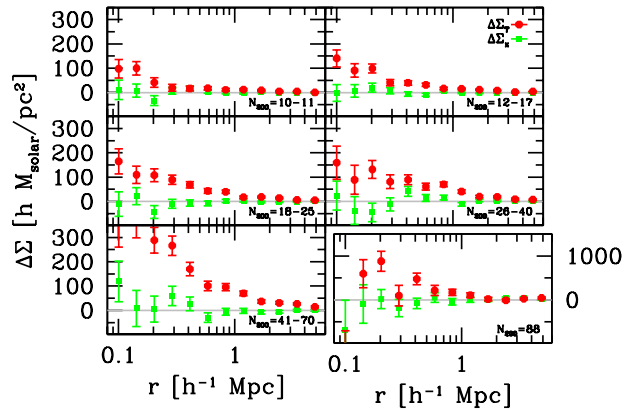}
  \caption{$\Delta\Sigma_t$ and $\Delta\Sigma_x$ profiles for six richness bins in one realization of the mock catalog.  Note the different scale in the $N_{200}=88$ bin.  A color version is available online.}
  \label{deltasigma-mock}
  \end{center}
\end{figure}

We test our analysis by creating a set of mock catalogs from our data.  We preserve the positions, shape errors, and photometric redshifts of the Stripe 82 galaxies and the positions, redshifts, and richnesses of the clusters, but replace the actual shears of the background galaxies with the expected shear generated from a shear model plus noise.  The input mass-richness relation is set to $M_{200|20}=9 \times 10^{13} h^{-1} M_{\sun}$ and $\alpha=1.4$.  We generate 20 realizations of the mock catalog of the base model: an NFW profile with Gaussian redshift errors, the expected geometry ($a=1$), and some of the haloes incorrectly centered on the BCG with a richness-dependent probability from simulations in \citet{Johnston}.  The shears are generated with an expectation value computed from the shear model and a random Gaussian error based on the variance of the observed shears with magnitude.  This error should contain both the intrinsic shape errors as well as measurement errors, and it asymptotes to 0.21 at $dered_i=18$.  The redshifts used to compute $\Sigma_c$ are drawn from a normal distribution using the photometric redshift error as the width.  A sample $\Delta\Sigma$ profile for one realization of a mock catalog is shown in Fig. \ref{deltasigma-mock}. In addition to these base model mock catalogs, we also generate several alternate cases: a singular isothermal sphere profile and an NFW model with $c_{200}=10$ to check the effect of incorrect profile assumptions; a true Gaussian redshift error that is ten times larger than the error reported in the catalog and used for analysis; and $\Omega_m=1$ rather than $0.3$.  We generate 5 realizations of each of the four alternate scenarios and analyze all the mock catalogs using the same pipeline we use for the data, including the assumption of an NFW model and photo-z errors as reported.    

\begin{figure}
  \begin{center}
  \includegraphics[width=0.45\textwidth]{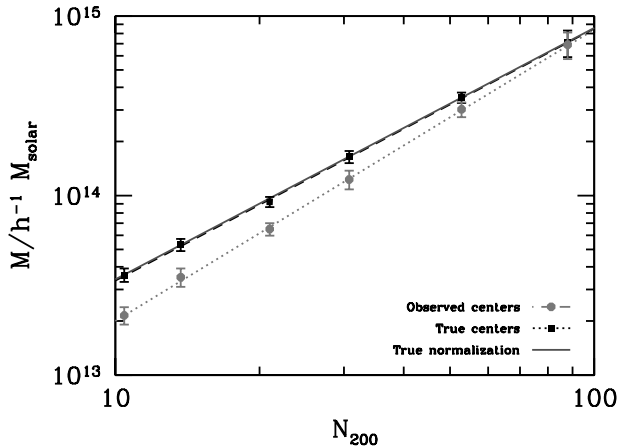}
  \caption{Mean and variance of the best-fit mass as a function of richness for twenty realizations of the mock catalog, analyzed using true centers and observed centers.}
  \label{mass-mocks}
  \end{center}
\end{figure}

\begin{figure}
  \begin{center}
  \includegraphics[width=0.45\textwidth]{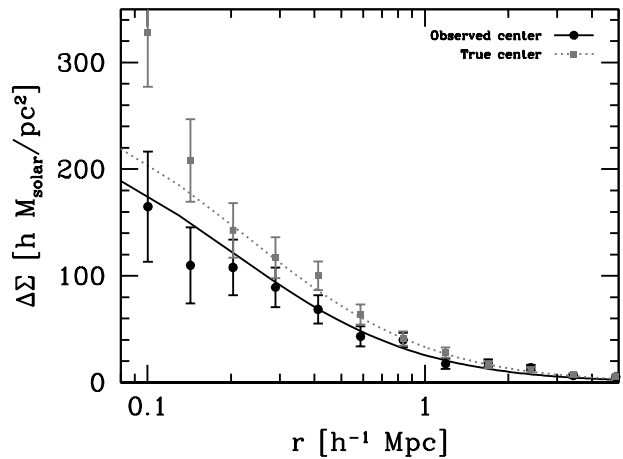}
  \caption{Sample $\Delta\Sigma$ profiles for the $N_{200}=$26-40 richness bin from one realization of the mock catalog.  Light points were analyzed with the correct centers and dark points were analyzed with the (offset) observed centers, with best-fit models for both.}
  \label{deltasigma-miscentering}
  \end{center}
\end{figure}

Next, we fit an NFW lens model to the mock data using a $\chi^2$ analysis with the mass as the only free parameter; we do not include the $C(R)$ misidentification correction as it was not used as an input to the mock catalogs.  The results from 20 base model cases are shown in Fig. \ref{mass-mocks}.  As expected, the mass is underestimated due to the dilution of the miscentered clusters: the mean best fit values of the twenty mock catalogs are $M_{200|20}=(6.17 \pm 0.32) \times 10^{13} h^{-1} M_{\sun}$ and $\alpha=1.62 \pm 0.08$. Using the true centers, we we fit $M_{200|20} =(8.89 \pm 0.34) \times 10^{13} h^{-1} M_\odot$ and $\alpha=1.40 \pm 0.06$, very close to the input model.  A sample $\Delta\Sigma$ profile using the two different centers for one bin in one realization of the mock catalog is shown in Fig. \ref{deltasigma-miscentering}.  Since the probability of miscentering decreases with increasing richness, the underestimation of the mass also decreases with increasing richness; we find
\begin{equation}
  \frac{M_{200,true}}{M_{200,mis}} = 1.44 \pm 0.17 \left( \frac{N_{200}}{20}\right)^{-0.21 \pm 0.18}
\end{equation}
which we will apply to the results of the Stripe 82 data as a correction.

\begin{figure}
  \begin{center}
  \includegraphics[width=0.45\textwidth]{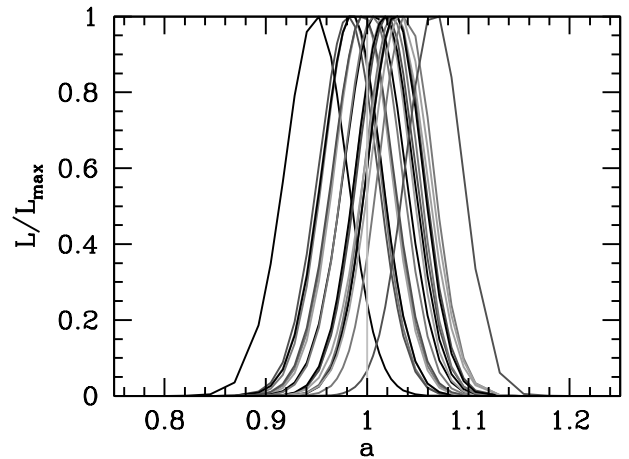}
  \caption{The likelihood of our tomography parameter $a$ in twenty realizations of the mock catalog.}
  \label{likelihood-mocks}
  \end{center}
\end{figure}

\begin{figure}
  \begin{center}
  \includegraphics[width=0.45\textwidth]{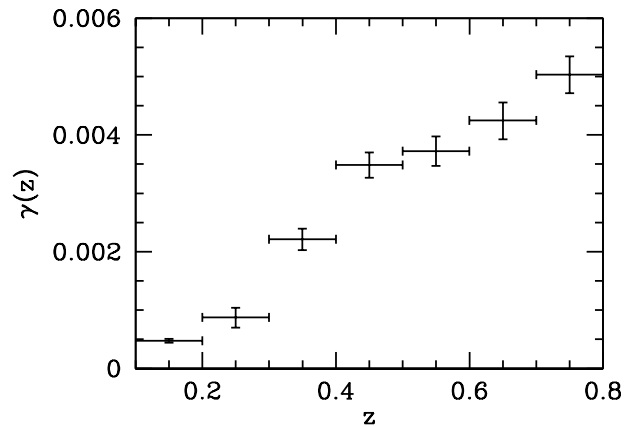}
  \caption{Shear signal as a function of background galaxy redshift for one realization of the mock catalog.  This is the shear at 1 $h^{-1}$ Mpc from a cluster of $N_{200}=20$ based on the fit to all clusters in the catalog.}
  \label{visualtomography-mocks}
  \end{center}
\end{figure}

The likelihood curves for $a$ for different realizations of the mock catalog are shown in Fig. \ref{likelihood-mocks}.  The 1$\sigma$ error obtained from each mock catalog is of order 0.03, and we find the peak position of the curves is clustered near our input value of 1 (mean$=1.01 \pm 0.007$).  We use the spread in the peaks to estimate the effects of the errors that exist but that we have not explicitly included in the fitting model, such as binning and photometric errors.  The errors as estimated from the standard deviation in the peak values of $a$ is $\sigma_{a} = 0.025$, similar to the width of an individual likelihood curve, as we expect.  We also show the results for the shear as a function of redshift fit for one realization of the mock catalog in Fig. \ref{visualtomography-mocks}.

\begin{figure} 
  \begin{center}
  \begin{tabular}{c}
  \includegraphics[width=0.47\textwidth]{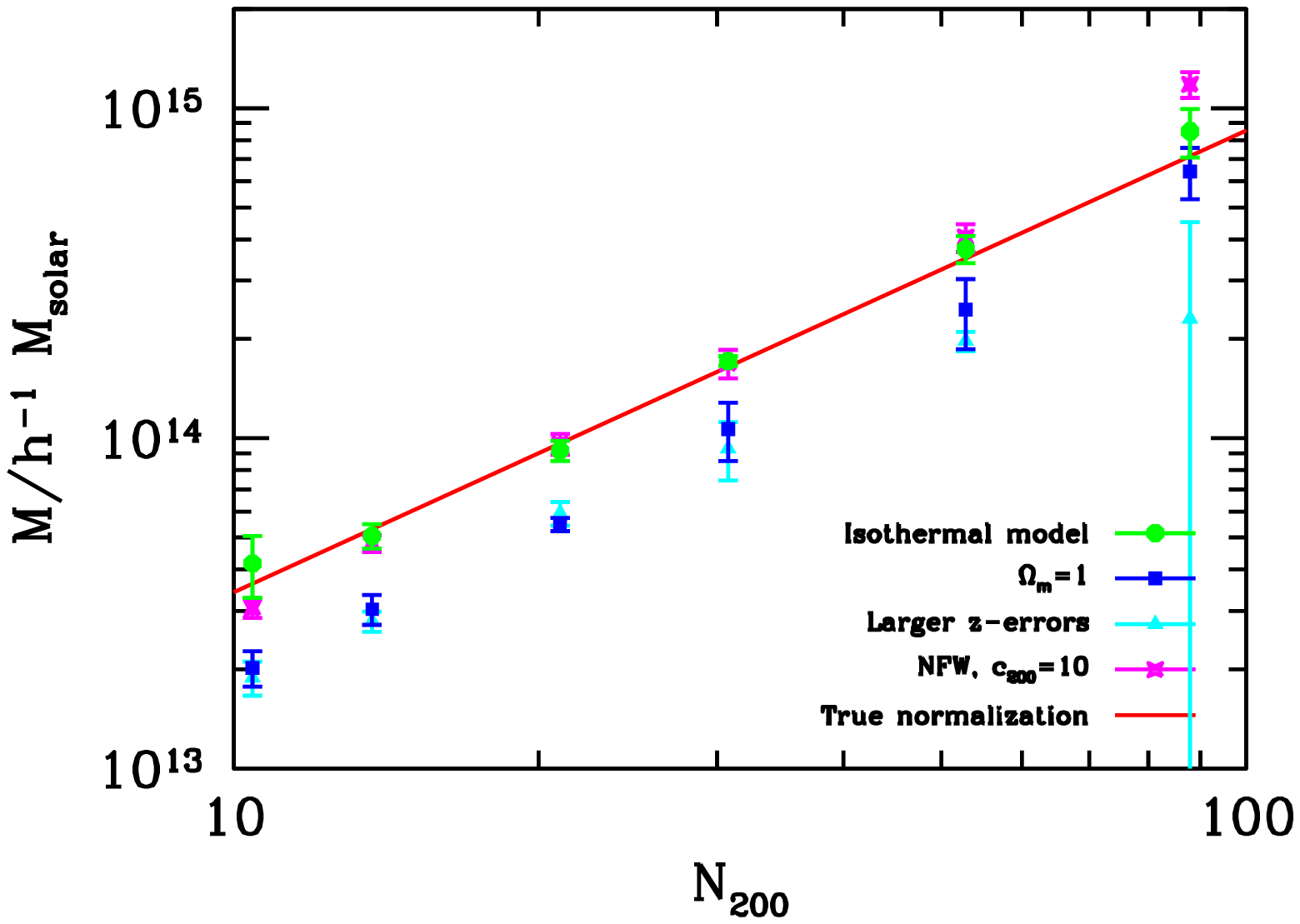} \\
  \includegraphics[width=0.45\textwidth]{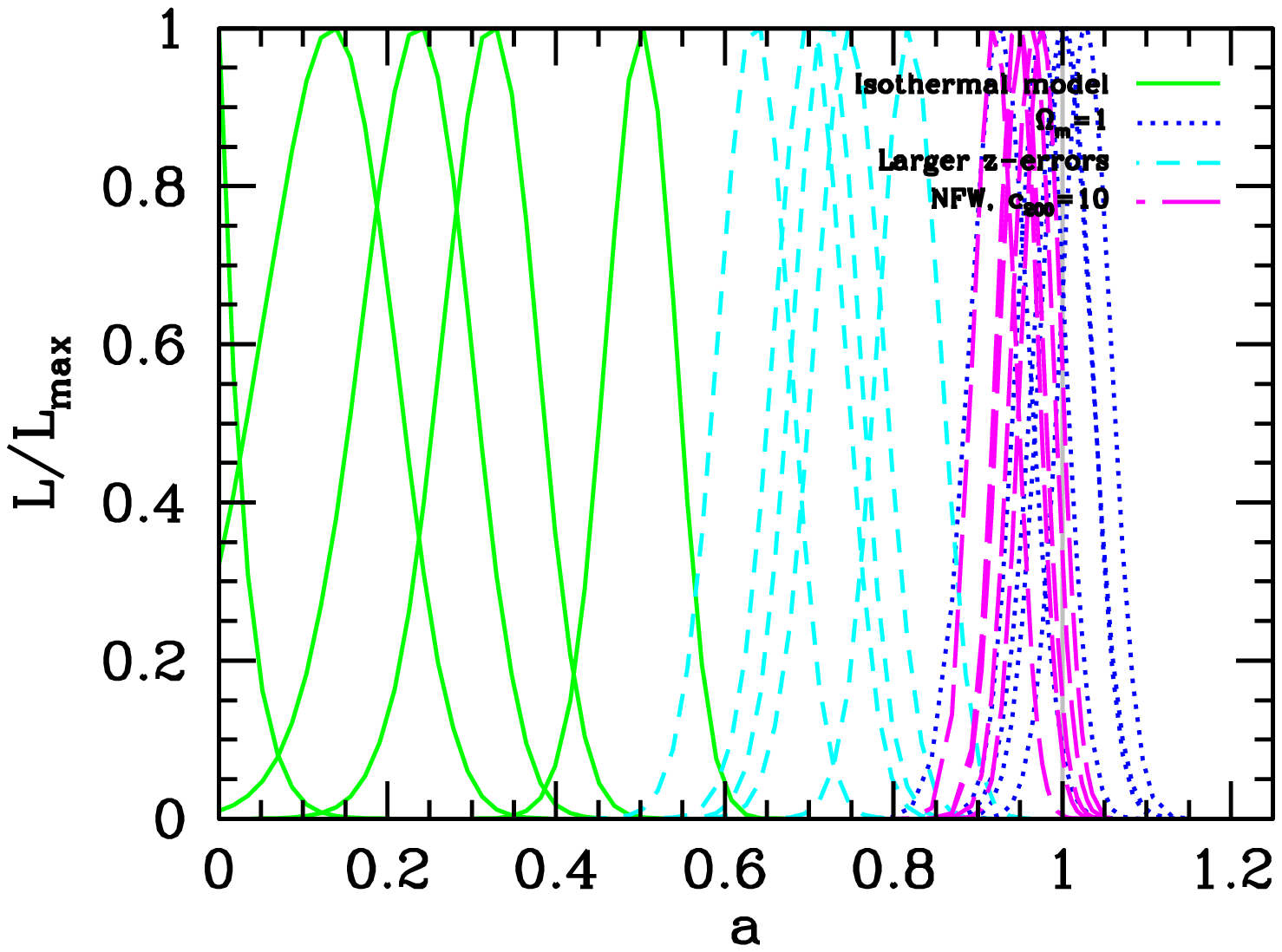}
  \end{tabular}
    \caption{Mass determinations and likelihood curves for four alternate mock data scenarios.  A color version is available online.}
  \label{alternate-mocks}
  \end{center}
\end{figure}

The results of the alternate scenarios are shown in Fig. \ref{alternate-mocks}.  We use the miscentered versions of these mocks to try to understand what the effect on our results would be from each of the errors described.  The isothermal case leads to overestimated masses and much too low peak likelihood values of $a$.  This means we should be able to distinguish the models by the use of tomography.  A similar, though much less dramatic, effect is seen with the NFW model with $c=10$: the peak likelihood value of $a$ is low, so again the tomography acts as a check on our profile assumptions.  The slope of the mass-richness relation is altered, but as we do not know the true value, this will not be distinguishable in the data.  Greater redshift errors lower the mass determinations, and they also significantly lower the peak values of the likelihood curves.  The altered cosmological model affects the mass but does not detectably alter the peak values of $a$, so we do not expect to be sensitive even to very different cosmological models.  

\section{Results}\label{Results}

\begin{figure}
  \begin{center}
  \includegraphics[width=0.5\textwidth]{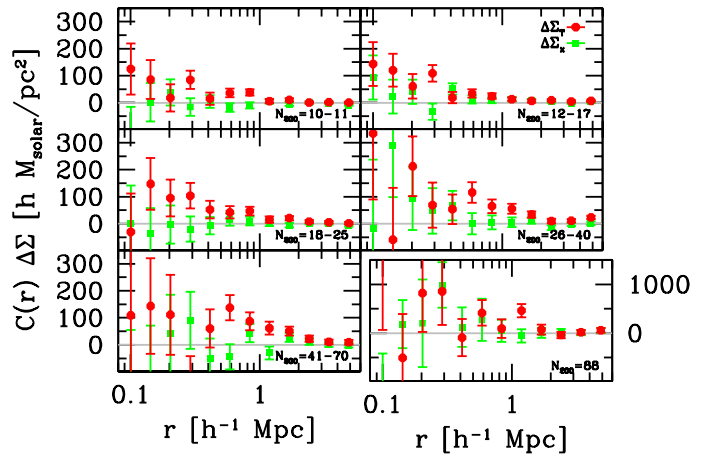}
  \caption{$\Delta\Sigma_t$ and $\Delta\Sigma_x$ profiles for six richness bins in the Stripe 82 coadd.  Note the different scale in the $N_{200}=88$ bin.  A color version is available online.}
  \label{deltasigma-data}
  \end{center}
\end{figure}

Fig.~\ref{deltasigma-data} shows the observed $\Delta\Sigma$, which removes most the dependence on cluster redshift so that clusters of the same mass have the same profile.  (The main dependence on $z$, the lensing geometry, is removed, but we retain a slight dependence due to the parameterization of the mass as proportional to $\rho_c$.)  As expected, the tangential signal $\Delta\Sigma_t$ is large close to the cluster centers and drops with radius, while the cross signal $\Delta\Sigma_x$ is consistent with no signal at all radii.  Also as expected, the signal is largest in the largest richness bins.

\begin{figure} 
  \begin{center}
  \includegraphics[width=0.5\textwidth]{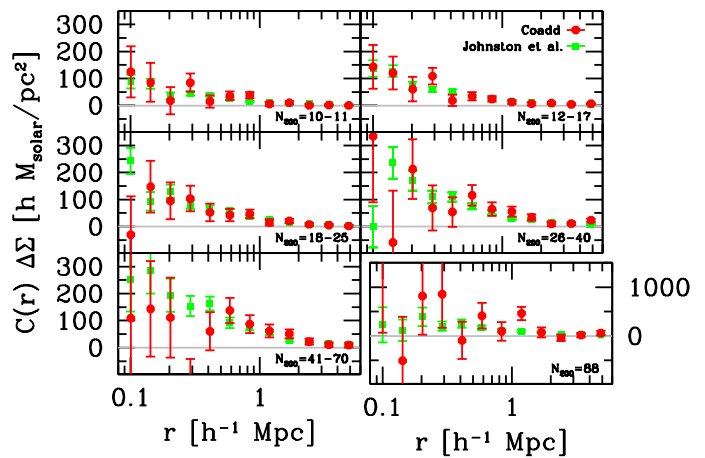}
  \caption{$\Delta\Sigma$ profiles for our analysis of Stripe 82 coadd and the results of \citet{Johnston} for a subset of DR4.  The signal increases close to the cluster centers, and also increases with richness, as expected.  The results are consistent with \citet{Johnston}.  Note the different scale in the $N_{200}=88$ bin.  A color version is available online.}
  \label{deltasigma-compare}
  \end{center}
\end{figure}

When we compare the $\Delta\Sigma$ results shown in Fig. \ref{deltasigma-compare} to the results of \citet{Johnston} for the entire SDSS catalog, we see that our amplitude is consistent, so we do not believe the coaddition process has diluted the lensing signal.

\begin{figure}
  \begin{center}
  \includegraphics[width=0.45\textwidth]{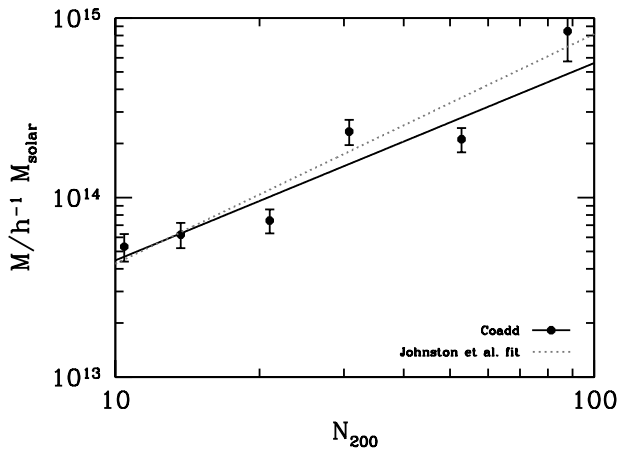}
  \caption{Best-fit masses as a function of richness for the Stripe 82 data.  We find a mass-richness relation of $M_{200}=(9.56 \pm 0.75) (N_{200}/20)^{1.10 \pm 0.12} \times 10^{13} h^{-1} M_{\sun}$.}
  \label{mass-data}
  \end{center}
\end{figure}

The mass-richness relation is shown in Fig. \ref{mass-data}.  This leads to best fit parameters $M_{200|20}=(9.56 \pm 0.75) \times 10^{13} h^{-1} M_{\sun}$ and $\alpha=1.10 \pm 0.12$. These mass estimates are also consistent with the \citeauthor{Johnston} results for the entire SDSS catalog (with the 18\% upward  correction of \citet{Rozo} due to photoz effects as described in \citet{Mandelbaum}).  We note again that our largest mass bin has only a single object, but combining this bin with the next lowest bin does not significantly change our results (changing $M_{200|20}$ by 1\% and $\alpha$ by 3\%) so we choose the binning that matches other analyses.

\begin{figure}
  \begin{center}
  \includegraphics[width=0.45\textwidth]{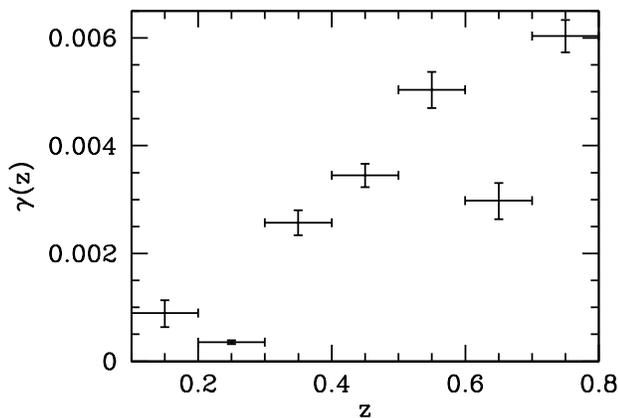}
  \caption{Shear signal as a function of background galaxy redshift for the Stripe 82 coadded data.  This is the shear at 1 $h^{-1}$ Mpc from a cluster of $N_{200}=20$ based on the fit to all clusters in the catalog.}
  \label{visualtomography-data}
  \end{center}
\end{figure}

A visual representation of the shear as a function of redshift is shown in Fig. \ref{visualtomography-data}.  We use the best-fit $\alpha$ from the mass-richness fit to compute the likelihood in $a$, which is shown in Fig. \ref{likelihood-data}.  We obtain a peak at $a=0.99$.  From the spread in the peak values of the mock data (Fig. \ref{likelihood-mocks}) this is consistent with the expected value of 1 and the mean mock data value of 1.01; adding the width of the likelihood curve and the dispersion of the peak values of the mock data in quadrature, we take the 1 $\sigma$ range in the data to be 0.94-1.02.  This results strongly disfavors the singular isothermal sphere as a model of the cluster profile.  Our 1$\sigma$ range is too large to draw conclusions about the concentration, however.  We also strictly rule out the case $a=0$, which corresponds to no redshift dependence in the shear signal.  


\section{Conclusion}

\begin{figure}[!h]
  \begin{center}
  \includegraphics[width=0.45\textwidth]{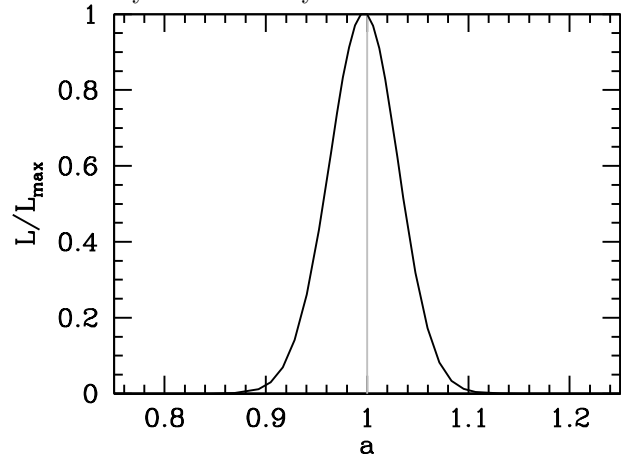}
  \caption{Likelihood for our tomography parameter $a$ generated using all the clusters in our Stripe 82 data set.  The value is close to 1, as expected.}
  \label{likelihood-data}
  \end{center}
\end{figure}

We have examined the lensing signal behind clusters in Stripe 82 of SDSS. The signal is consistent with the wider, shallower Data Release 4, as evidenced by Fig.~\ref{mass-data}, which shows the amplitude and slope of the mass-richness relation. The deeper sample considered here, supplemented with photometric redshifts, allowed us to measure the effect that galaxies further from the cluster are sheared more than those nearby. Fig.~\ref{likelihood-data} illustrates the constraints on a parameter $a$ that encodes this effect. The estimate of $a=0.99\pm0.04$ represents a clean detection of this tomographic signal.  In addition, the detection of $a$ requires an accurate model of the lensing profile to marginalize over, and we find that the singular isothermal sphere is not a sufficient description of the masses of the clusters in this analysis.

The lensing signal does not appear to be systematically corrupted by the coadd process, even though this was not driven by lensing requirements. This bodes well for future surveys that will rely on coadded data.

\section{Acknowledgements}

This work is supported by the US Department of Energy, including grant DE-FG02095ER40896; and by National Science Foundation Grant AST-0908072.  The authors would like to thank Rachel Mandelbaum and the anonymous referee for helpful comments.


\begin{thebibliography}{99}
\bibliographystyle{plainnat}

\bibitem[Abazajian et al.(2009)]{coadd} 
   Abazajian, K.~N., Adelman-McCarthy, J.~K., Ag\"{u}eros, M.~A., et al. 2009, \apjs, 182, 543

\bibitem[Adelman-McCarthy et al.(2008)]{DR6}
   Adelman-McCarthy, J.~K., Ag\"{u}eros, M.~A., Allam, S.~S., et al. 2008, \apjs, 175, 27

\bibitem[Albrecht et al.(2006)]{TaskForce}
    Albrecht, A., Bernstein, G., Cahn, R., et al. 2006, arXiv:astro-ph/0609591

\bibitem[Annis et al.(2011)]{Annis}
   Annis, J., Soares-Santos, M., Strauss, M., et al. 2011, arXiv:1111.6619
	
\bibitem[Bernstein(2007)]{Bernstein}
   Bernstein, G. 2007, in ASP Conf. Ser. 371, Statistical Challenges in Modern Astronomy IV, ed. G. Jogesh Babu \& Eric D. Feigelson, 59

\bibitem[Drinkwater et al.(2010)]{WiggleZ}
  Drinkwater, M.~J., Jurek, R.J., Blake, C., et al. 2010, \mnras, 401, 1429

\bibitem[Frieman et al.(2008)]{Stripe82}
   Frieman, J.~A., Bassett, B., Becker, A., et al. 2008, \aj, 135, 338

\bibitem[Gao et al.(2008)]{Concentration}
  Gao, L., Navarro, J.~F., Cole, S., et al. 2008, \mnras, 387, 536
   	
\bibitem[Garilli et al.(2008)]{VVDS}
  Garilli, B., Le F\`{e}vre, O., Guzzo, L., et al. 2008, \aap, 486, 683
  
\bibitem[Gavazzi \& Soucail(2007)]{Gavazzi}
  Gavazzi, R., \& Soucail, G. 2007, A\&A, 462, 459
  
\bibitem[Hirata \& Seljak(2003)]{HirataSeljak}
  Hirata, C. \& Seljak, U. 2003, \mnras, 343, 459

\bibitem[Hoekstra \& Jain(2008)]{HoekstraJain}
   Hoekstra, H. \& Jain, B. 2008, Annual Review of Nuclear and Particle Science, 58, 99

\bibitem[Hu(1999)]{Hu}
  Hu, W. 1999, \apjl, 522, L21

\bibitem[Huterer(2010)]{Huterer}
  Huterer, D. 2010, General Relativity and Gravitation, 42, 2177

\bibitem[Johnston et al.(2007)]{Johnston}
   Johnston, D.~E., Sheldon, E.~S., Wechsler, R.~H., et al. 2007, arXiv:0709.1159

\bibitem[Kaiser \& Squires(1993)]{KaiserSquires}
   Kaiser, N. \& Squires, G. 1993, \apj, 404, 441

\bibitem[Koester et al.(2007)]{Koester}
   Koester, B.~P., McKay, T.~A., Annis, J., et al. 2007, \apj, 660, 239 
   
\bibitem[Lin et al.(2011)]{cosmicshear}
   Lin, H., Dodelson, S., Seo, H-J., et al. 2011, arXiv:1111.6622
    
\bibitem[Mandelbaum et al.(2005)]{magnification}
  Mandelbaum, R., Hirata, C.~M, Seljak, U., et al. 2005, \mnras, 361, 1287
    
\bibitem[Mandelbaum et al.(2008)]{Mandelbaum}
   Mandelbaum, R., Seljak, U., Hirata, C.~M., et al. 2008, \mnras, 386, 781
   
\bibitem[Marian et al.(2009)]{Marian}
   Marian, L., Smith, R.~E., \& Bernstein, G.~M. 2009, \apjl, 698, L33

\bibitem[Medezinski et al.(2011)]{Medezinski}
   Medezinski, E., Broadhurst, T., Umetsu, K., Ben\'{i}tez, N., and Taylor, A. 2011, 414, 1840

\bibitem[Mellier(1999)]{Mellier}
   Mellier, Y. 1999, \araa, 37, 127

\bibitem[Munshi et al.(2008)]{Munshi}
   Munshi, D., Valageas, P., van Waerbeke, L., \& Heavens, A. 2008, \physrep, 462, 67
   
\bibitem[Navarro, Frenk \& White(1997)]{NFW}
   Navarro, J.~F., Frenk, C.~S. and White, S.~D.~M. 1997, \apj, 490, 493

\bibitem[Oyaizu et al.(2008)]{DR6-photoz}
  Oyaizu, H., Lima, M., Cunha, C.~E., et al. 2008, \apj, 674, 768

\bibitem[Peacock et al.(2006)]{Peacock}
   Peacock, J.~A., Schneider, P., Efstathiou, G., et al. 2006, arXiv:astro-ph/0610906
	
\bibitem[Reis et al.(2011)]{photoz}
   Reis, R.~R.~R., Soares-Santos, M., Annis, J., et al. 2011, \apj, submitted, arXiv:1111.6620

\bibitem[Rozo et al.(2010)]{Rozo}
  {Rozo}, E., {Wechsler}, R.~H., {Rykoff}, E.~S., et al. 2010, \apj, 708, 645
	
\bibitem[Schneider et al.(2006)]{SchneiderBook}
   Schneider, P., Kochanek, C.~S., \& Wambsganss, J. 2006, 33rd Saas-Fee Advanced Course, Gravitational Lensing: Strong, Weak, and Micro (Berlin: Springer)

\bibitem[Shan et al.(2011)]{Shan}
  Shan, H., Kneib, J.-P., Tao, C., et al. 2011, arXiv:1108.1981   
   
\bibitem[Sheldon et al.(2004)]{Sheldon}
   Sheldon, E.~S., Johnston, D.~E., Frieman, J. A., et al. 2004, \aj, 127, 2544
   
\bibitem[Sheldon et al.(2009)]{SheldonStripe82}
   Sheldon, E.~S., Johnston, D.~E., Scranton, R., et al. 2009, \apj, 703, 2217
   
\bibitem[Simon et al.(2011)]{Simon}
   Simon, P., Heymans, C., Schrabback, T., et al. 2011, arXiv:1109.0932 

\bibitem[Taylor et al.(2004)]{Taylor}
   Taylor, A.~N., Bacon, D.~J., Gray, M.~E., et al. 2004, \mnras, 353, 1176

\bibitem[Wang \& Steinhardt(1998)]{Wang}
   Wang, L. \& Steinhardt, P.~J. 1998, \apj, 508, 483

\bibitem[Weiner et al.(2005)]{DEEP2}
  Weiner, B.~J., Phillips, A.~C., Faber, S.~M., et al. 2005, \apj, 620, 59

\bibitem[Wright \& Brainerd(2000)]{Wright}
   Wright, C.~O. \& Brainerd, T.~G. 2000, \apj, 534, 34     

\bibitem[Yee et al.(2000)]{CNOC2}
   Yee, H.~K.~C., Morris, S.~L., Lin, H., et al. 2000, \apjs, 12, 475

\bibitem[York et al.(2000)]{SDSS}
   York, D.~G., Adelman, J., Anderson, J.~E., et al. 2000, \aj, 120, 1579
   
\bibitem[Zhao et al.(2009)]{Zhao}
   Zhao, G.-B., Pogosian, L., Silvestri, A., \& Zylberberg, J. 2009, \prl, 103, 24   

\end{thebibliography}
\end{document}